\documentclass[reprint,
aps,
pra,
]{revtex4-2}

\usepackage{graphicx}
\usepackage{dcolumn}
\usepackage{bm}
\usepackage{color}

\begin{document}

\title{Polarization effects in cosmic-ray acceleration by cyclotron autoresonance}

\author{Yousef I. Salamin}
\email{ysalamin@aus.edu}
\affiliation{Department of Physics, American University of Sharjah, POB 26666, Sharjah, United Arab Emirates}

\begin{abstract}
Employing a two-parameter model for representing the radiation field, the theory of cosmic-ray acceleration by cyclotron autoresonance is analytically generalized here to include any state of polarization. The equations are derived rigorously and used to investigate the dynamics of the nuclides $_1$H$^1$, $_2$He$^4$, $_{26}$Fe$^{56}$, and $_{28}$Ni$^{62}$, in severe astrophysical conditions. Single-particle calculations and many-particle simulations show that these nuclides can reach ZeV energies ($1 ~ZeV  = 10^{21}$ eV) due to interaction with superintense radiation of wavelengths $\lambda=1~$ and $10~ \mu$m, and $\lambda=50$ pm, and magnetic fields of strengths at the mega- and gigatesla levels. Examples employing radiation intensities in the range $10^{32}-10^{42}$ W/m$^2$ are discussed.
\end{abstract}

\maketitle

\section{Introduction}

This work is part of theoretical efforts dedicated to exploring features of the recently advanced mechanism of cyclotron autoresonance acceleration (CARA) as a possible explanation for the ZeV energies of ultrahigh-energy cosmic rays (UHECR). For about fifty years now, cosmic-ray particles carrying energies in excess of 10$^{18}$ eV have been seen by a number of detectors around the world. A substantial body of work, experimental and theoretical, has been dedicated to pinpointing their origin and explaining their enormous energies \cite{harari,PhysRevLett.10.146,aab2017inferences,abbasi2,aharonian3,allard,bell,drury,drury2,ANCHORDOQUI20191,arcavi2017optical,PhysRevD.97.063010,bertaina2019search,PhysRevLett.125.121106,PhysRevD.102.062005}. CARA has recently been advanced \cite{Salamin_2021,SALAMIN2021127275} as a possible mechanism of acceleration for protons and heavier nuclei to ZeV energies, in an environment where radiation of wavelengths at the ends of a wide range coexist with megatesla (MT) to gigatesla (GT) uniform magnetic fields \cite{price,rosswog,sukhbold,Frederiks_2013}.  

Here, the role that could be played in CARA by polarization of the radiation field is investigated \cite{Granot_2003,capparelli,sagar2012exact,sagar2014effect}. 
Issues regarding plausibility of the classical approach, radiation loss, and the radiation-reaction effects, will all be left out as they have been discussed elsewhere \cite{Salamin_2021}. Examples taken up in the applications will be representative of the observed cosmic-ray spectrum. Cosmic-ray particles are close to 90\% protons, roughly 10\% are alpha particles, and the rest are nuclei of heavier atoms.

The basic theory will be laid out in the next section, employing a two-parameter representation for the general state of the elliptically-polarized radiation field. In our most recent work \cite{SALAMIN2021127275}, the polarization effects were also the subject of detailed discussion, albeit employing a single-parameter model, with identical conclusions.  Parametric equations for a single-particle trajectory and exit kinetic energy, during interaction with the combined electromagnetic fields will be derived in the next section. In Sec. \ref{sec:app}, acceleration of single particles will be investigated by the mechanism when left-elliptically polarized radiation is used. It will be shown that optimal acceleration takes place in the presence of a left circularly-polarized (LCP) radiation field. Illustrative examples, and detailed investigations, will focus on the acceleration of a proton, an alpha particle, and the nuclei of iron and nickel, using LCP fields. In those examples, radiation fields of wavelengths 50 pm and $1~\mu$m, and intensities $10^{42}$ and $10^{37}$ W/m$^2$, respectively, will be used. Further examples employing, what may arguably be considered as, a more realistic set of parameters will be presented and discussed in Sec. \ref{sec:real}. Due to the fact that the boundaries which separate neighboring parts of the electromagnetic spectrum are not sharp in general, wavelength of $\lambda=50$~ pm will be referred to, admittedly loosely in this work, as representative of a GRB. Likewise, radiation of wavelength $\lambda=1~\mu$m will be referred to, equally as loosely, as {\it visible} light. A summary of the work and our main conclusions will be given in Sec. \ref{sec:sum}.

\section{Theory}\label{sec:theo}

\begin{figure}
	\centering
	\includegraphics[width=8cm]{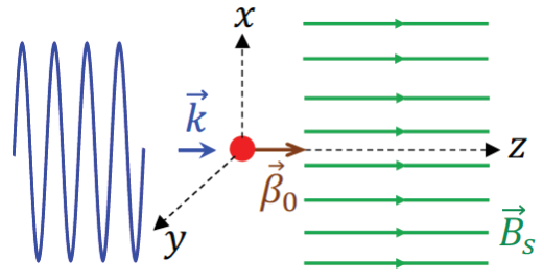}
	\caption{A schematic diagram showing the conditions for single-particle acceleration by cyclotron autoresonance. The radiation wavefront has been assumed to catch up with the particle at $t=0$, the instant it passes through the origin of a conveniently chosen Cartesian coordinate system.}
	\label{fig1}
\end{figure}

Figure \ref{fig1} is a schematic diagram showing a particle of mass $M$ and charge $Q$, injected with velocity $\bm{\beta}_0=\beta_0\hat{\bm{k}}$, scaled by $c$, the speed of light in vacuum, parallel to the lines of a uniform magnetic field $\bm{B}_s = B_s\hat{\bm{k}}$.  Shown also is a plane-wave elliptically-polarized pulse of radiation, propagating along the same direction as the particle. The wavefront of the pulse will be assumed to catch up with the particle at a point ($x_0, y_0, z_0$) of the Cartesian coordinate system, precisely at $t=0$ \cite{hartemann}.

Implicit in this scheme is the assumption that the particle is pre-accelerated inside the progenitor by some other mechanism, such as shock waves produced by a supernova explosion \cite{soker2017magnetar} or a binary neutron-star merger \cite{Abbott_2017}. Subsequently, the particle is assumed to encounter the topology shown schematically in Fig. \ref{fig1}. As such, CARA may be thought of as a booster acceleration scheme.

The radiation field will be modeled by an elliptically-polarized plane-wave of frequency $\omega$ and wavevector  $\bm{k} = (\omega/c)\hat{\bm{k}}$. Thus, the combined electric and magnetic fields in the problem may now be written as \cite{jackson}
\begin{eqnarray}	
	\label{E} \bm{E}(\eta) &=& E_0\left[\hat{\bm{i}} \cos\eta+\hat{\bm{j}} A \cos(\eta+\phi)\right],\\
	\label{B} \bm{B}(\eta) &=& \frac{E_0}{c}\left[-\hat{\bm{i}} A \cos(\eta+\phi) + \hat{\bm{j}}\cos\eta\right]+\hat{\bm{k}}B_s,
\end{eqnarray}
in which $E_0$ is a constant amplitude, $\eta=\omega t-kz$ is the phase of the radiation field, and $\hat{\bm{i}}, \hat{\bm{j}}$, and $\hat{\bm{k}}$ are unit vectors along the positive $x-$, $y-$, and $z-$axes, respectively, of the Cartesian coordinate system of Fig. \ref{fig1}. Also, $A$ and $\phi$ are the polarization parameters. For example, the case of $A=0$ represents linear polarization and the states for which $A=1$ and $\phi=\pm\pi/2$ describe circular polarization, and so on.

The radiation field intensity, defined as the magnitude of the time-averaged Poynting vector, will be needed to determine $E_0$ for a given intensity. Using the radiation parts of Eqs. (\ref{E}) and (\ref{B}) yields
\begin{equation}\label{I}
	I = \frac{1}{2} c\varepsilon_0E_0^2 (1+A^2),
\end{equation}
where $\varepsilon_0$ is the permittivity of free space. Intensities to be used in the applications below are in the range $10^{37}-10^{42}$ W/m$^2$. Radiation of such intensity may be associated with compact objects or a binary neutron-star merger \cite{Frederiks_2013,kann2019highly}. Let the luminosity of some event be $\sim10^{48}$ W. If only half of the output energy \cite{fan2006interpretation,Abbott_2017} of an event like this is radiant and is beamed \cite{Frail_2001,Granot_2003,RevModPhys.76.1143} through a circle of radius 100 m, the radiation in this case can have an intensity $I\sim10^{43}$ W/m$^2$. An alternative calculation assuming the radiant energy is emitted isotropically, as if by a point source as opposed to being beamed as has recently been observed \cite{Abbott_2017} following the event GW170817, would give an intensity many orders of magnitude smaller than this. From a point (or a spherically-symmetric) source, the intensity falls off with the square of the distance from the center of the emitter. This issue will be revisited in Sec.  \ref{sec:real} below.

Employing the phase $\eta$ as a variable, in place of the time $t$, the relativistic equations of motion of the particle in the above fields, namely (SI unit are used throughout)
\begin{eqnarray}
\label{mom}\frac{d\bm{p}}{dt} &=& Q[\bm{E}+c\bm{\beta}\times\bm{B}],\\
\label{en}\frac{d{\cal E}}{dt} &=& Qc\bm{\beta}\cdot\bm{E},
\end{eqnarray}
admit solutions in closed analytic form \cite{PhysRevA.62.053809,PhysRevA.58.3221,PhysRevA.61.043801}. In Eqs. (\ref{mom}) and (\ref{en}) the relativistic momentum and energy are $\bm{p}=\gamma Mc\bm{\beta}$ and ${\cal E}=\gamma Mc^2$, respectively, where $\bm{\beta}$ is the particle's velocity in units of $c$, and $\gamma=(1-\beta^2)^{-1/2}$. In terms of the radiation-field phase, the initial conditions on the position translate into $\eta_0 = -kz_0$.
In developing the analytic solutions, a resonance condition is arrived at, which may be expressed by \cite{PhysRevA.58.3221,PhysRevA.61.043801,Salamin_2021,sagar2014effect}
\begin{equation}
	\label{r} r = \frac{QB_s}{M\omega}\sqrt{\frac{1+\beta_0}{1-\beta_0}} = 1.
\end{equation}
Resonance actually amounts to the cyclotron frequency of the particle around the lines of the magnetic field, $\omega_c=QB_s/M$, matching the Doppler-shifted frequency $\omega_D=\omega\sqrt{(1-\beta_0)/(1+\beta_0)}$ of the radiation field, which the particle senses in its own rest frame. This allows for the rate of change of the particle's relativistic energy $d{\cal E}/dt=Qc\bm{\beta}\cdot\bm{E}$ to stay positive, guaranteeing continuous energy buildup or acceleration. In the absence of resonance, this rate oscillates in sign, which results in energy gain from interaction with one-half of every radiation-field cycle encountered and loss during interaction with the subsequent half cycle.

Of central importance for our investigations in the present publication are the parametric equations giving the trajectory of a particle, as well as its Lorentz factor (energy scaled by $Mc^2$, the rest energy). Without loss of generality, the initial position will be taken at the origin of the Cartesian coordinate system; see Fig. \ref{fig1} for a schematic. In terms of the radiation field phase, this amounts to having $\eta_0=0$. The equations in question may be arrived at following along the lines of earlier work \cite{PhysRevA.58.3221,PhysRevA.61.043801,Salamin_2021}. When the resonance condition is used, the trajectory and scaled energy equations take the following forms

\begin{eqnarray}
\label{x} x(\eta) &=& \left(\frac{ca_0}{2\omega}\right)\gamma_0(1+\beta_0)\left[X_0+X_1A\right],\\
\label{y} y(\eta) &=& \left(\frac{ca_0}{2\omega}\right) \gamma_0(1+\beta_0)\left[Y_0+Y_1A\right],\\
\label{z} z(\eta) &=& \frac{c}{\omega}\left(\frac{1+\beta_0}{1-\beta_0)}\right)\left\{\left(\frac{\beta_0}{1+\beta_0}\right)\eta\right.\nonumber\\
& &\left.+\left(\frac{a_0^2}{96}\right)\left[Z_0+Z_1A+Z_2A^2\right]\right\},\\
\label{g} \gamma(\eta) &=& \gamma_0+\left( \frac{a_0^2}{16}\right) \gamma_0(1+\beta_0)\left[\Gamma_0+\Gamma_1A+\Gamma_2A^2\right],
\end{eqnarray}
where 
\begin{eqnarray}
X_0 &=& \eta \sin \eta,\\
X_1 &=& \sin (\eta +\phi)-\eta  \cos(\eta+\phi)+(\cos\eta-2) \sin\phi,\\
Y_0 &=& \eta \cos\eta-\sin\eta,\\
Y_1 &=& \eta  \sin(\eta+\phi)-\sin\eta\sin\phi,\\
Z_0 &=& 4\eta \left(\eta^2+3\sin^2\eta\right),\\
Z_1 &=& 2\left(4\eta^3-6\eta+3 \sin2 \eta\right) \sin\phi,\\
Z_2 &=& 4 \eta^3+6\eta -6\eta\cos(2\eta +2\phi)-3\sin 2\eta\nonumber\\
& &+3 \sin(2\eta +2\phi)-3\left(2 \eta^2+1\right)\sin2\phi,\\
\Gamma_0 &=& 2\eta^2+1-\cos2\eta+2 \eta\sin2\eta,\\
\Gamma_1 &=& 2\left(2\eta^2+\cos2\eta-1\right) \sin\phi,\\
\Gamma_2 &=& 2\eta^2-\cos2\eta+1+4\eta\sin\eta \cos
(\eta+2\phi).
\end{eqnarray}
The above equations will be used to investigate the dynamics, mainly issues related to the evolution of the kinetic energy, of a single particle under CARA conditions.

\section{Applications}\label{sec:app}

 Examples of particles to accelerate will be nuclei of hydrogen, helium, iron and nickel. In particular, the nuclides $_1$H$^1$, $_{2}$He$^{4}$,  $_{26}$Fe$^{56}$, and $_{28}$Ni$^{62}$ will be chosen as representative of the whole spectrum of the known cosmic rays. Recall that cosmic rays are mostly protons and alpha particles. Furthermore, $_{26}$Fe$^{56}$ and $_{28}$Ni$^{62}$ are known to be the most stable nuclei in nature, which qualifies them as candidates for acceleration under the severe conditions of CARA, while remaining intact during the acceleration process. As such, they are also good candidates for surviving decay by photo-pion production and other processes in intergalactic space \cite{abbasi1,abbasi2,murase2008high,fang2012newly,zhang2018low}.

\begin{figure}[t]
	\includegraphics[width=8cm]{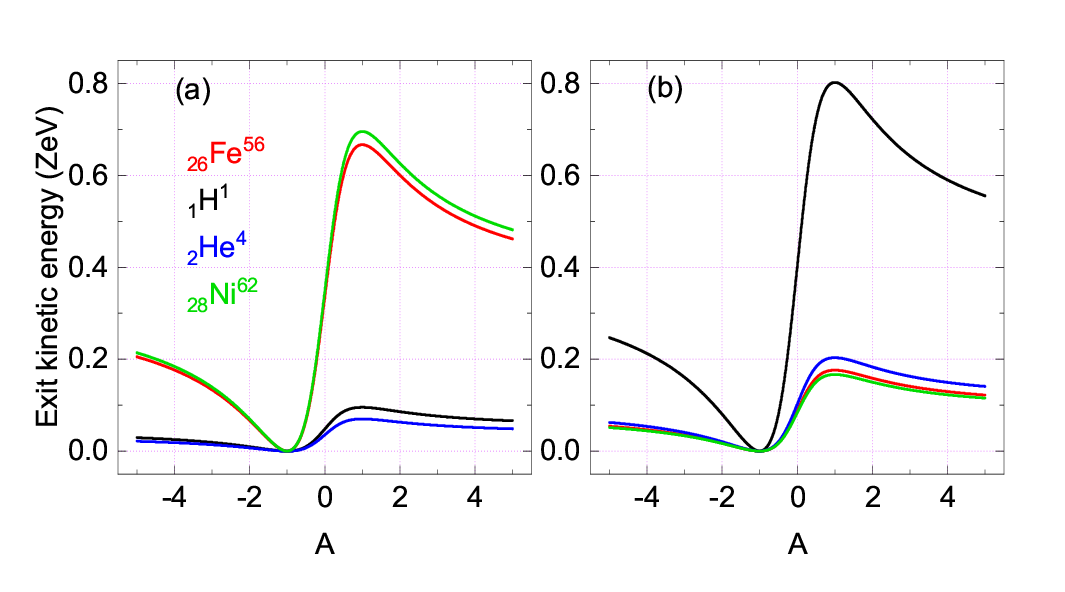}
	\caption{Dependence of the exit kinetic energy on the polarization parameter, $A$, of the nuclides $_1$H$^1$, $_2$He$^4$, $_{26}$Fe$^{56}$, and $_{28}$Ni$^{62}$. {\it Exit} here means end of interaction time equivalent to 5 phase cycles of the radiation field.  (a) Initial scaled speed is derived from the initial kinetic energy $K_0 =  200$ MeV, radiation field intensity is $I=10^{37}$ W/m$^2$, and wavelength is $\lambda=1~ \mu$m. (b) Initial scaled speed is derived from the initial kinetic energy $K_0 =  30$ TeV, radiation-field intensity is $I=10^{42}$ W/m$^2$, and its wavelength is $\lambda=50~ $pm. For each case, $B_s$ follows from Eq. (\ref{r}).  }
	\label{fig2}
\end{figure}

In addition to the fact that all nuclides get accelerated to ZeV energy, employing the given parameter sets, three more observations cannot be missed in Fig. \ref{fig2}. (a) The case of $A=1$ and $\phi=\pi/2$, obviously one of left circular polarization, results in maximum exit kinetic energy $K_m$. This is the case in which the electric field vector of the plane-wave rotates about the direction of propagation in a left-handed sense. If the particle also gyrates around the lines of $\bm{B}_s$ in a left-handed sense, then its velocity vector maintains a fixed, almost zero, angle with the electric field. This is autoresonance, a condition under which the rate of change of the particle's energy $d{\cal E}/dt=Q\bm{v}\cdot\bm{E}>0$, all the time. This means the energy of the particle will increase monotonously, leading to a maximum exit kinetic energy. The value $A_m=1$ for which $K_m$ may be attained, can easily be obtained from setting $\partial\gamma(\eta,A)/\partial A=0$ and using $\eta=2n\pi$, where $n$ is an integer. (b) The case $A=-1$ and $\phi=\pi/2$ results in zero exit kinetic energy (zero gain). To explain this, note that the fields of $A=-1$ and $\phi=\pi/2$ are the same as those of $A=+1$ and $\phi=-\pi/2$, or right circular-polarization (RCP). For this situation, the electric field rotates about the direction of propagation in a right-handed sense, opposite that of the particle gyration. This causes the sign of $d{\cal E}/dt$ to flip at the end of interaction with every half-cycle of the radiation field. The result of that sign oscillation is that energy gained during one-half of a field-cycle gets lost completely during interaction with the immediately following half, i.e., zero net gain. (c) The case corresponding to $A=0$ is one of linear polarization (LP). The exit kinetic energy in this case is one-half that of interaction with an LCP wave of the same amplitude. This can be appreciated by considering an LP plane-wave as composed of an LCP and an RCP of the same amplitude. All the energy gained by the particle comes from the LCP component, while interaction with the RCP wave results in no energy gain, as has just been explained. 

In anticipation of the important result that gain of energy by a particle from the radiation field is a maximum when a left circularly-polarized (LCP) radiation field is employed, left elliptically-polarized radiation will be used first. For such radiation, $\phi=\pi/2$ and $A$ can be arbitrary. Exit kinetic energies of the four nuclides are shown in Fig. \ref{fig2}, at the end of interaction with 5 radiation-field phase cycles, as a function of the polarization parameter $A$. Two sets of curves are presented, one for interaction with visible light and a second with the fields of a gamma-ray burst (GRB). Injection into the interaction region follows the schematic diagram of Fig. \ref{fig1}, and the parameters of the scheme are all given in the figure caption \cite{kann2019highly}.

\begin{figure}[t]
	\centering
	\includegraphics[width=8cm]{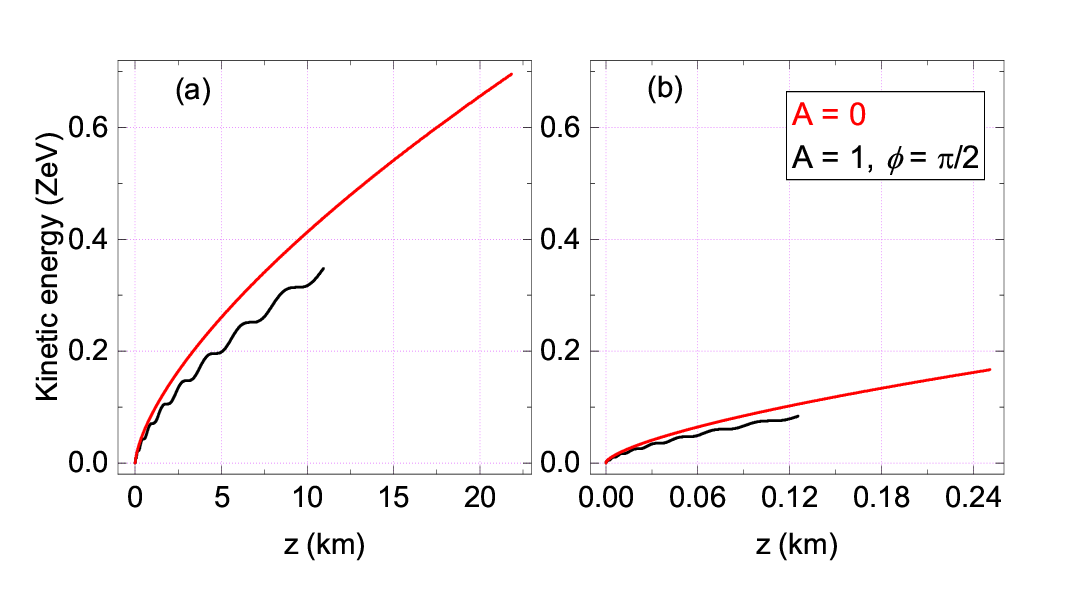}
	\caption{Variations of the kinetic energy with the axial excursion distance, during an interaction time interval equivalent to 5 phase cycles of the radiation field, of the nuclide $_{28}$Ni$^{62}$. The particle's initial position is at ($0, 0, 0$) in all cases. (a) Initial scaled speed is derived from the initial kinetic energy $K_0 =  200$ MeV, visible radiation-field intensity is $I=10^{37}$ W/m$^2$ and wavelength is $\lambda=1~ \mu$m, and $B_s\simeq39.7298$ MT. (b) Initial scaled speed is derived from the initial kinetic energy $K_0 = 30$ TeV, GRB radiation-field intensity is $I=10^{42}$ W/m$^2$ and its wavelength is $\lambda=50~$pm, and $B_s\simeq828.677$ MT. }
	\label{fig3}
\end{figure}

Figure \ref{fig3} presents examples of acceleration by CARA to ZeV energies of the nuclide $_{28}$Ni$^{62}$. Preacceleration  \cite{murase2008high,fang2012newly,zhang2018low} to MeV and TeV energies is assumed prior to submission to  ultraintense visible light and GRB radiation fields, respectively, together with superstrong magnetic fields. In both cases, the gain from the LP radiation is clearly one-half that from an LCP wave, as expected. The figure shows that interaction with the visible radiation results in acceleration to higher energy than with the much more intense GRB. There is little basis for comparison here as the cases correspond to two widely differing parameter sets. Furthermore, the acceleration efficiency in the latter case is much higher than in the former. This can be seen by comparing the energy gradients, roughly obtained from dividing the exit kinetic energies by the corresponding total excursion distances. For example, the energy gradient from the LCP wave is $\sim31.847$ EeV/m (visible) and $\sim664.255$ EeV/m (GRB). Here, EeV stands for exaelectronvolt (1 EeV $= 10^{18}$ eV).

Closer scrutiny reveals that the particle gains energy from the radiation field at the highest rate  during interaction with a small part of the first radiation-field phase cycle it encounters. The time  rate at which the kinetic energy of the particle changes during interaction with the radiation field follows from Eq. (\ref{g}) with the help of the constant of the motion \cite{Salamin_2021} \begin{equation}
    \gamma_0(1-\beta_0)=\gamma(1-\beta_z),
\end{equation} 
and employing 
\begin{equation}
    \frac{d}{dt}=\frac{d\eta}{dt} \frac{d}{d\eta}=\omega(1-\beta_z)\frac{d}{d\eta}.
\end{equation} 
The result may finally  be written as
\begin{equation}
\label{dK} \frac{dK}{dt} = I\lambda^2 \xi(\eta).
\end{equation}
Equation (\ref{dK}) gives an expression for the power delivered to the particle by the radiation field. In this equation,
\begin{equation}
\label{xi} \xi(\eta) = \frac{Q^2}{2\pi\epsilon_0 Mc^2\lambda(A^2+1)}\left[\frac{\xi_0+\xi_1A+\xi_2A^2}{\gamma}\right],
\end{equation}
where
\begin{eqnarray}
\xi_0 &=& \cos\eta
	[\sin\eta+\eta  \cos\eta],\\
\xi_1 &=& \sin\phi \left[\eta -\sin\eta \cos\eta\right],\\
\xi_2 &=& \cos(\eta+\phi) \left[\eta  \cos(\eta+\phi)+\sin\eta \cos\phi\right].
\end{eqnarray}
The quantity $\xi(\eta)$ is actually a dimensionless factor which may be thought of as an acceleration efficiency parameter, a measure of the efficiency of energy conversion from electromagnetic (in the radiation field) to kinetic (of the particle). 

\begin{figure}
	\centering
	\includegraphics[width=4cm,height=4cm]{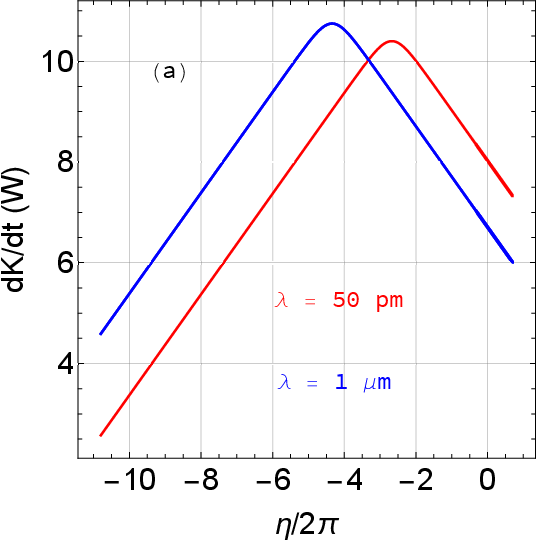}
	\includegraphics[width=4cm,height=4cm]{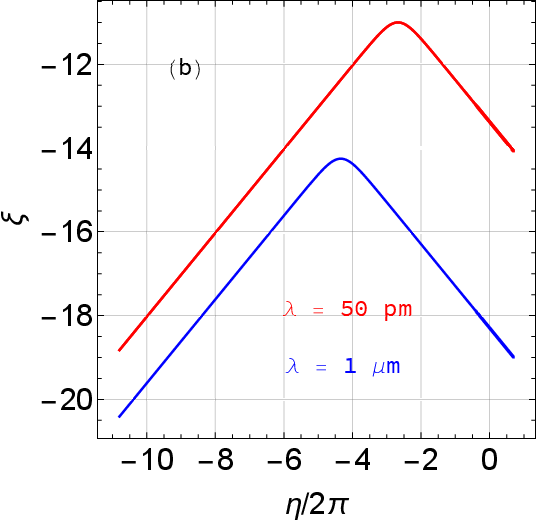}
	\caption{Log-Log plots of evolution with the number of radiation-field phase cycles of: (a) the time-rate of kinetic energy change, and (b) the acceleration efficiency parameter, for the nuclide $_{28}$Ni$^{62}$, accelerated by cyclotron autoresonance. In all cases, the initial position is ($0, 0, 0$). Fields of left circularly-polarized radiation ($A=1$ and $\phi=\pi/2$) are employed.  In the case of $\lambda = 1 ~\mu$m, the initial scaled speed is derived from the initial kinetic energy $K_0 =  200$ MeV,  the intensity is $I=10^{37}$ W/m$^2$, and the magnetic field strength is $B_s\simeq39.7298$ MT. When the GRB fields are used, the parameters are: initial scaled speed derived from $K_0 = 30$ TeV, intensity $I=10^{42}$ W/m$^2$, wavelength $\lambda=50~ $pm, and $B_s\simeq828.677$ MT.}
	\label{fig4}
\end{figure}

Examples of the energy conversion rate and acceleration efficiency parameter are shown in Fig. \ref{fig4} for the nuclide $_{28}$Ni$^{62}$ during interaction with LCP visible light and a GRB. The figure clearly shows a positive rate that is quite sizable throughout the interaction period. It also goes through a spike during interaction with a small part of the first phase cycle it encounters. recall that acceleration of the particle causes it to lose energy by the emission of radiation. This leads to a decrease in the energy conversion rate (and acceleration efficiency) with time. The acceleration gets slowed down also by the radiation-reaction effects, which act in a way reminiscent of the macroscopic force of friction, to retard the motion of the particle. These issues have been briefly discussed elsewhere \cite{Salamin_2021}. 

\section{More realistic  parameters}\label{sec:real}

Observation of a binary neutron-start merger, source of the gravitational waves detected recently \cite{abbott}, showed that the event was followed by emission of ultraintense radiation, with frequencies spanning the entire electromagnetic spectrum \cite{Drout1570,Hallinan1579,Kilpatrick1583,Wu_2019}. Gamma-rays, in the form of a gamma ray burst (GRB), x-rays, and visible light were all beamed out in opposite directions from the polar caps of the newly formed compact object. Among other things, these emissions are a clear indicator of stellar nucleosynthesis and the presence of atoms \cite{pian2017spectroscopic,arcavi2017optical,smartt2017kilonova,kajino2019current,wang2020}. A simple calculation, based on the assumption that most of the output energy was radiant and beamed through a circle of radius $\sim100$ m, results in the GRB intensities exceeding $10^{42}$ W/m$^2$, for example, that have been employed in \cite{Salamin_2021} and in this work, thus far.
It may be argued that the parameters employed in our calculations are unrealistic. At the heart of the counter-argument against adoption of  these intensity values is the assumption that energy is radiated isotropically (not beamed) in which case the calculated intensities would be many orders-of-magnitude lower.  

In a recent followup, many-particle calculations have been performed, in which the assumption has been made that the particle-particle interactions could be neglected. This is justified by the mere fact that the number densities employed are very small compared to those in a typical solid, where such interactions can not be ignored. 
Furthermore, the focus so far has been on examples already considered in \cite{Salamin_2021}, with the aim of lending support to the single-particle results by performing simulations for non-interacting many particles. Not only do the many-particle simulations agree, in general, with the single-particle calculations, but they do not seem to depend on the size of the ensemble employed. 

\begin{figure}
	\centering
	\includegraphics[width=7cm]{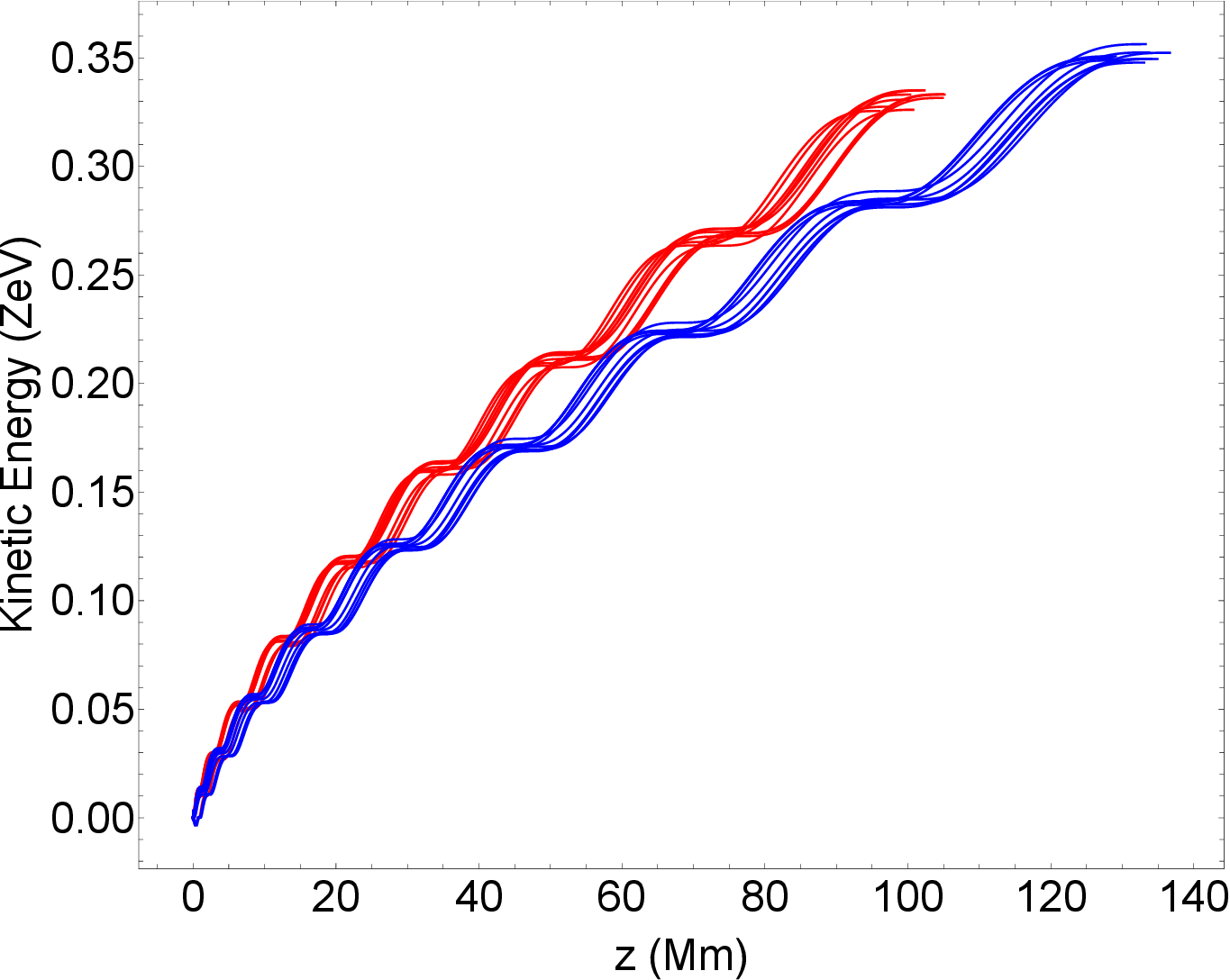}
	\caption{Kinetic energy evolution with the excursion distance (in $10^6$ m) of an ensemble of nuclei of Fe$^{+26}$ (blue) and Ni$^{+28}$ (red) employing the fields of infrared light of wavelength $\lambda = 10 ~ \mu$m and intensity $I=10^{32}$~ W/m$^2$. The initial ensemble has $N = 10$ nuclei inside a disk of radius 5 m and thickness $H=\lambda$ (number density $n_d \simeq 12732$ m$^{-3}$). Initial ensemble kinetic energy: normal distribution of mean $\bar{K}_0=30$ TeV and standard deviation $\Delta K_0=0.3$ TeV. The average resonance magnetic field strengths sensed by the ensemble members and calculated on the basis of Eq. (\ref{r}) are $B_s \simeq (3640\pm 24)$ T (iron) and $(4145\pm32)$ T (nickel).}
	\label{fig5}
\end{figure}

The above points will be addressed together by adoption of a more realistic parameter set. Key departure from the old parameter set is achieved by employing infrared radiation of wavelength $10~ \mu$m and intensity $I=10^{32}$ W/m$^2$. An ensemble of only 10 nuclei is employed, making the particle density $n_d\simeq12732$ m$^{-3}$ and, therefore, strengthening the argument in support of the particle-particle interactions being negligibly small. Simulations have been performed for Fe$^{+26}$ and Ni$^{+28}$, two of the most stable nuclei in nature, and the results are displayed graphically in Fig. \ref{fig5}.

Figure \ref{fig5} shows evolution of the kinetic energies of the particles of ensembles of nuclei of iron (blue) and nickel (red) during interaction with 5 phase cycles of the radiation field. All other ensemble parameters are given in the figure caption. The exit kinetic energies are as follows:  $K_e\simeq(0.3507\pm0.0024$) ZeV, for iron, and $K_e\simeq(0.3324\pm0.0033$) ZeV, for nickel. Note that the difference between the two sets of results is quite small. This is due to the fact that the masses and charge-to-mass ratios of the two nuclei are quite close. Note also that the resonance magnetic field  strengths are $B_s \sim(3640\pm24)$ T (iron) and $B_s \sim(4145\pm32)$  T (nickel). These numbers are more than two-orders-of-magnitude smaller than the previous ones presented in Sec. \ref{sec:app}.

\section{Inter-particle effects}\label{pp}

The investigations thus far have been single-particle or have involved non-interacting many particles. This section is devoted to inclusion of the particle-particle interaction effects.

To account for the effect on the motion of particle number $j$ due to the fields generated by all the other particles in the ensemble, the equations of motion (\ref{mom}) and (\ref{en}) will now be replaced by
\begin{eqnarray}
	\label{momj}\frac{d\bm{p}_j}{dt} &=& Q[\bm{E}_j+\bm{E}_j^{pp}+c\bm{\beta}_j\times\left(\bm{B}_j+\bm{B}_j^{pp}\right)],\\
	\label{enj}\frac{d{\cal E}_j}{dt} &=& Qc\bm{\beta}_j\cdot\left(\bm{E}_j+\bm{E}_j^{pp}\right).
\end{eqnarray}
Here, $\bm{E}_j(\eta)$ and $\bm{B}_j(\eta)$ are given by Eqs. (\ref{E}) and (\ref{B}) and $\bm{E}_j^{pp}$ and $\bm{B}_j^{pp}$ stand for the particle-particle interaction fields.

Contributions to the electric and magnetic fields at the position of the $j^{th}$ particle due to all the other particles in the ensemble, follow from the Li\'enard-Wiechert potentials. The electric field contribution is \cite{jackson}
\begin{eqnarray}
	\bm{E}_j^{pp}(t) &=& \frac{Q}{4\pi\epsilon_0}\sum_{i\neq j}\left[\frac{\hat{\bm{n}}_{ij}-\bm{\beta}_i}{\gamma_i^2(1-\hat{\bm{n}}_{ij}\cdot\bm{\beta}_i)^3R_{ij}^2}\right.\nonumber\\
	& &\left. +\frac{\hat{\bm{n}}_{ij}\times(\hat{\bm{n}}_{ij}-\bm{\beta}_i)\times{\bm{a}}_i}{c(1-\hat{\bm{n}}_{ij}\cdot\bm{\beta}_i)^3R_{ij}}\right]_{ret} ,
\end{eqnarray}
where $\bm{R}_{ij}\equiv \bm{r}_j-\bm{r}_i$ is position vector of the $j^{th}$ particle relative to that of the $i^{th}$ particle, $\hat{\bm{n}}_{ij}=\bm{R}_{ij}/R_{ij}$ is a unit vector pointing from the $i^{th}$ particle to the $j^{th}$ particle, $\bm{\beta}_i$ denotes velocity of the $i^{th}$ particle scaled by the speed of light, $\gamma_i=(1-\beta_i^2)^{-1/2}$, and $\bm{a}_i\equiv d\bm{\beta}_i/dt$ is the acceleration scaled by $c$. Furthermore, $ret$ indicates that the quantities inside the square brackets are to be evaluated at the retarded time $t_r\equiv t-R_{ij}/c$. 

The corresponding magnetic field contribution, on the other hand, is given by
\begin{equation}
	\bm{B}_j^{pp}(t)  = \frac{1}{c}\hat{\bm{n}}_{ij}\times\bm{E}_j^{pp}(t) 
\end{equation}
When the interaction terms are properly included, the resulting equations of motion (\ref{momj}) and (\ref{enj}) can no longer be solved analytically, which also implies violation of the resonance condition (\ref{r}). In what follows, Eqs. (\ref{momj}) and (\ref{enj}) will be solved numerically, employing the radiation and magnetic fields obtained under resonance, with the interaction terms introduced as corrections.

\section{Summary and conclusions}\label{sec:sum}

The main working equations for a particle, accelerated by cyclotron autoresonance, have been derived employing elliptically-polarized radiation. Without referring to a specific astrophysical environment \cite{VINK2008503,kaspi,soker2017magnetar,smartt2017kilonova} known with certainty to host the required conditions, examples of single-particle acceleration have been discussed, in which the particles are representative of the main parts of the cosmic-ray mass spectrum. In particular, acceleration to ZeV energies has been demonstrated for protons, alpha particles, and the most stable nuclides in nature, namely, those of iron and nickel. All results presented in this work support the conclusion that ZeV kinetic energies can be reached by cosmic-ray particles via the cyclotron autoresonance acceleration (CARA) mechanism. 

It has also been shown that the particle's kinetic energy attains a maximum value, $K_m$, when the mechanism employs left circularly-polarized radiation. Under the same conditions, the particle reaches a kinetic energy of half that much from interaction with a linearly-polarized wave of the same amplitude.

In laser acceleration, the particles gain energy by {\it simultaneous} multi-photon absorption from a coherent source. In the zevatron, the radiation is absorbed from a source that may or may be coherent \cite{Gainullin,kunihito,benedetti}. Incoherent absorption of radiation in the zevatron may involve an extremely short delay between successive absorption events. In each separate event, a photon catches up with, and gets absorbed by, a particle that is already moving at an ultra-relativistic speed. A photon generated at the progenitor a short while later catches up with, and gets absorbed by, the same particle after a short delay, and so on. Thus, the tremendous energy gained by the particle from the radiation field can come from absorption of a tremendous number of photons, made available to it simultaneously from a coherent source, or at some extremely short delays from an incoherent one.

CARA has been proposed as a booster accelerator scheme \cite{Salamin_2021}. The assumption has been made all along that a particle is typically pre-accelerated to MeV-TeV energies inside the source, by other mechanisms described by the shock-wave model, for example, before submission to CARA for further acceleration. The pre-acceleration models take into detailed consideration the complex plasma background inside the source.

This work is supported by Faculty Research Grant AS1811 (American University of Sharjah).

\bibliographystyle{apsrev4-2}

\bibliography{Zevatron}

\end{document}